\documentclass[showpacs,aps,prd,nofootinbib,floatfix,amsmath,amssymb]{revtex4}
\usepackage{graphicx}
\usepackage{relsize}
\begin{document}

\makeatletter
%Feynman slash
\newbox\slashbox \setbox\slashbox=\hbox{$/$}
\newbox\Slashbox \setbox\Slashbox=\hbox{\large$/$}
\def\pFMslash#1{\setbox\@tempboxa=\hbox{$#1$}
  \@tempdima=0.5\wd\slashbox \advance\@tempdima 0.5\wd\@tempboxa
  \copy\slashbox \kern-\@tempdima \box\@tempboxa}
\def\pFMSlash#1{\setbox\@tempboxa=\hbox{$#1$}
  \@tempdima=0.5\wd\Slashbox \advance\@tempdima 0.5\wd\@tempboxa
  \copy\Slashbox \kern-\@tempdima \box\@tempboxa}
\def\FMslash{\protect\pFMslash}
\def\FMSlash{\protect\pFMSlash}
\def\miss#1{\ifmmode{/\mkern-11mu #1}\else{${/\mkern-11mu #1}$}\fi}
%%%% Uso:  \pFMSlash{p}
\makeatother

%\tightenlines
\title{Decays $Z\to gg\gamma$ and $Z'\to gg\gamma$ in the minimal $331$ model}
\author{A. Flores-Tlalpa, J. Monta\~no, F. Ram\'\i rez-Zavaleta, and J. J. Toscano}
\address{Facultad de Ciencias F\'{\i}sico Matem\'aticas,
Benem\'erita Universidad Aut\'onoma de Puebla, Apartado Postal
1152, Puebla, Puebla, M\'exico.}
\begin{abstract}
The one-loop induced $Z\to gg\gamma $ and $Z'\to gg\gamma $ decays are studied within the context of the minimal $331$ model, which predicts the existence of new gauge bosons and three exotic quarks. It is found that the $Z\to gg\gamma$ decay is insensitive to the presence of the exotic quarks, as it is essentially governed by the first two families of known quarks. As to the $Z'\to gg\gamma$ decay, it is found that the exotic quark contribution dominates and that for a heavy $Z'$ boson it leads to a $\Gamma(Z'\to gg\gamma)$ that is more than one order of magnitude larger than that associated with $\Gamma(Z'\to ggg)$.
\end{abstract}


\maketitle

In a recent communication~\cite{OP}, we presented a comprehensive study on the $Z\to ggg$ and $Z'\to ggg$~\cite{PTT} decays within the context of the minimal $331$ model~\cite{331-1,331-2}. This model, which is based in the $SU_C(3)\times SU_L(3)\times U_X(1)$ gauge group, has the peculiarity of predicting new physics effects at the TeV scale~\cite{331-2, Ng331}. In the gauge sector, the model predicts the existence of a new $Z'$ gauge boson as well as two doubly charged and two simply charged gauge bosons, that arise in the first stage of spontaneous symmetry breaking (SSB) when $SU_C(3)\times SU_L(3)\times U_X(1)$ is broken into $SU_C(3)\times SU_L(2)\times U_Y(1)$~\cite{TT}. These charged gauge bosons, which arise as a doublet of the $SU_L(2)$ group~\cite{TT}, are known as bileptons because they carry two units of lepton number. In this model, the lepton spectrum is the same\footnote{ It should be mentioned that there is a different version of this model~\cite{PT} which introduces exotic leptons but with the same quark sector.} as in the standard model (SM), but three new exotic quarks are needed in order to cancel anomalies~\cite{331-2,TT,Zp331}. To endow with mass the spectrum of particles, the model requires of a Higgs sector composed of three triplets and one sextet of $SU_L(3)$~\cite{Ng331,TT}. Interestingly, the new gauge boson masses are bounded from above~\cite{331-2,Ng331,Zp331} due to the theoretical constraint which yields $\sin^2\theta_W\equiv s^2_W\leqslant 1/4$~\cite{331-2,Ng331}. The fact that the value of $s^2_W$ is very close to $1/4$ at the $m_{Z'}$ scale leads to an upper bound on the scale associated with the first stage of SSB, which translates directly into a bound on the $Z'$ mass given by $m_{Z'}\leqslant 3.1$ TeV~\cite{Ng331}, which in turns implies that the bilepton masses cannot be heavier than $m_{Z'}/2$~\cite{Ng331}. It turns out that when $s^2_W(\mu)=1/4$ the coupling constant $g_X$ associated with the $U_X(1)$ group becomes infinite and a Landau pole arises~\cite{LP}.

\begin{figure}
\centering
\includegraphics[width=2.0in]{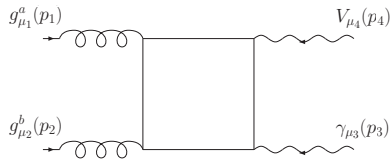}
\caption{\label{FD} Generic box diagram contributing to the $V\gamma gg$ vertex $(V=Z,Z')$. There are a total of 6 box diagrams, which are obtained from the one given in this figure via Bose symmetry.}
\end{figure}

In this brief report, we present results for the $Z\to gg\gamma$ and $Z'\to gg\gamma$ decays within the context of the $331$ model. These decays receive contributions from both the known quarks and the exotic quarks ($D$ and $S$ with charge $-4/3$ in units of the positron charge, and $T$ with charge $+5/3$). It turns out that the amplitude for the $Vgg\gamma$ vertex, with $V=Z,Z'$, can be obtained as a particular case of the amplitude for the $Vggg$ vertex, whose exact expression is listed in Ref.~\cite{OP}. As it is pointed out in Refs.~\cite{OP,Bij}, the amplitude for the $Vggg$ vertex is composed by the vector amplitude and the axial vector amplitude, which are finite and gauge invariant by themselves and do not interfere among themselves, as they are proportional to the color structures $d_{abc}$ and $f_{abc}$, respectively. However, due to the Abelian character of the photon, there is no contribution of axial vector type to the $Vgg\gamma$ vertex, the contribution is only of vector type and arises from box diagrams of the type shown in Fig.~\ref{FD}. Using the notation and conventions shown in this figure, the amplitude for the $V\to gg\gamma$ decay can be written as
\begin{equation}
{\cal M}_{V\to gg\gamma}={\cal M}^{\mu_1 \mu_2 \mu_3 \mu_4}_{ab}\epsilon^{*\;a}_{\mu_1}(p_1,\lambda_1)\epsilon^{*\;b}_{\mu_2}(p_2,\lambda_2)\epsilon^*_{\mu_3}(p_3,\lambda_3)\epsilon_{\mu_4}(p_4,\lambda_4),
\end{equation}
where
\begin{equation}
{\cal M}^{\mu_1 \mu_2 \mu_3 \mu_4}_{ab}=g^{q}_{VV}Q_{q}\delta_{ab}\Big(-\frac{ig^2_seg_VN_C}{2\pi^2}\Big)\sum_{j=1}^{18}f_{V_j}^qT^{\mu_1\mu_2\mu_3\mu_4}_{V_j},
\end{equation}
  with $Q_{q}$ representing the electric charge in units of the positron charge. The form factors $f_{V_j}^q$ and the gauge structures $T^{\mu_1\mu_2\mu_3\mu_4}_{V_j}$ appearing in the above expression are given in Ref.~\cite{OP}. As it is shown in this reference, these form factors are free of ultraviolet divergences, whereas the gauge structures satisfy the transversality conditions:
\begin{equation}
p_{i\mu_i}{\cal M}^{\mu_1 \mu_2 \mu_3 \mu_4}_{ab}=0\; , \; i=1,2,3,4.
\end{equation}
On the other hand, the decay width can be written as follows:
\begin{align}
\Gamma(V\to gg\gamma)=\,&\frac{m_V}{1536\,\pi^3}\int^1_0\int^1_{1-x}|{\cal M}|^2dxdy \nonumber \\
=\,&\frac{\alpha^2_s(m_V)\alpha^2N^2_Cm_V}{32\,\pi^3s^2_Wc^2_W}8\int^1_0\int^1_{1-x}
\sum^{18}_{q,q'}g^{q}_{VV}Q_{q}g^{q'}_{VV}Q_{q'}\Bigg(\frac{1}{3}
\sum_{\lambda_1,\lambda_2,\lambda_3,\lambda_4}{\cal V}_{q}{\cal V}^*_{q'}\Bigg)dxdy,
\end{align}
where
\begin{equation}
{\cal V}_q=\sum_{j=1}^{18}f_{V_j}^qT^{\mu_1\mu_2\mu_3\mu_4}_{V_j}\epsilon^{a*}_{\mu_1}(p_1,\lambda_1)
\epsilon^{b*}_{\mu_2}(p_2,\lambda_2)\epsilon^*_{\mu_3}(p_3,\lambda_3)\epsilon_{\mu_4}(p_4,\lambda_4)\; .
\end{equation}

We now turn to discuss our results. As already discussed in Ref.~\cite{OP}, the behavior of the amplitude associated with this class of four vector couplings is of decoupling nature when considered as a function of the internal mass quark, but it is nondecoupling when evaluated with respect to the mass difference of the members of the doublets of the SM quarks. This behavior was discussed in detail in Ref.~\cite{OP} and we will refrain from presenting here. Since the only difference between the vector amplitudes for the $Vgg\gamma$ and $Vggg$ couplings comes from coupling constants, color factors, and statistical factors, roughly one can to expect that the decay width for the $V\to gg\gamma$ transition is suppressed with respect to the one associated with the $V\to ggg$ process by a factor of $(36/5)(\alpha/\alpha_s)Q^2_{q_i}\approx 0.14\;Q^2_{q_i}$, which is always lower than the unity still for exotic quarks. Although this is the case for the $Z\to gg\gamma$ decay, we will see below that the exotic quark contribution to the $Z'\to gg\gamma$ transition leads to a width decay which can be one order of magnitude or larger than that associated with the $Z'\to ggg$ transition. The reason for this behavior is due to a constructive interference effect among the exotic quarks.

The $Z\to gg\gamma$ decay has already been calculated in the SM~\cite{ZSM}. Here, we present complete numerical results in the context of the minimal $331$ model, which include the contributions induced by the exotic quarks, for whose masses we will use values that are consistent with current bounds estimated in the literature~\cite{EQMass}. Following Ref.~\cite{OP}, we write the decay width of the $V\to gg\gamma$ transition as the sum of three partial widths:
\begin{equation}
\Gamma (V\to gg\gamma)=\Gamma_{q_i}+\Gamma_{Q_i}+\Gamma_{q_i-Q_i}\; ,
\end{equation}%%%%%%%%%%%%%%%%%%%%%%%%%%%%%%%%%%%%%%%%%%%%%%%%%%%%%%%%%%%%%%%%%%%%%%%%%%%%%%%%%%%%%%%%%%%%%%%
where $\Gamma_{q_i}$, $\Gamma_{Q_i}$, and $\Gamma_{q_i-Q_i}$ are the contributions of the SM quarks, the exotic quarks, and the interference between these contributions, respectively. In Table \ref{TABLE1}, the contributions to the decay width of the $Z\to gg\gamma$ transition induced by the three known families and the exotic quarks, which are computed with a degenerate mass of $m_{Q_i}=500$ GeV, are shown. From this table, it can be appreciated that the decay width is determined essentially by the first two families and a constructive interference effect. As it can be appreciated from this table, the contribution induced by the exotic quark is completely marginal. Using the value for the total $Z$ decay width reported by The Particle Data Group~\cite{PDG}, one obtains the following branching ratio
\begin{equation}
\mathrm{Br}(Z\to gg\gamma)=5.49\times 10^{-6}\;.
\end{equation}

\begin{table}[htb]
\caption{\label{TABLE1} Family contribution to the $\Gamma(Z\to gg\gamma)$ decay in the standard model.  Here $\Gamma^I_{q_i}$ represents the interference effect among families. The exotic quark contributions are also indicated.}
\begin{ruledtabular}
\begin{center}
\begin{tabular}{|c|r|c|c|c|c|}
  Family & $\Gamma$ [GeV] & $\Gamma^I$ [GeV] & $\Gamma_{q_i}$ [GeV] & $\Gamma_{Q_i}$ [GeV] & $\Gamma_{q_i-Q_i}$ [GeV]\\
  \hline
  $u,d$ & $2.34\times10^{-6}$ & - & - & - & - \\
 \hline
  $c,s$ & $2.4\times10^{-6}$ & - & - & - & - \\
  \hline
  $t,b$ & $5.81\times10^{-7}$ & - & - & - & - \\
  \hline
  Total & $5.32\times10^{-6}$ & $8.36\times10^{-6}$ & $1.37\times10^{-5}$ & $\sim 10^{-11}$ & $\sim -10^{-10}$
\end{tabular}
\end{center}
\end{ruledtabular}
\end{table}

\begin{figure}
\centering
\includegraphics[width=3.1in]{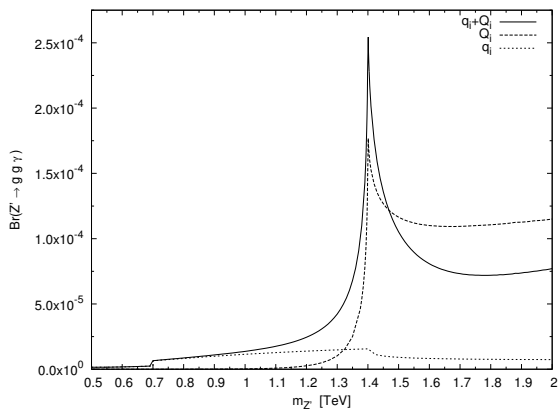}
\caption{\label{Graphic} Impact of the known and exotic quarks on $\mathrm{Br}(Z'\to gg\gamma)$ as a function of the $Z'$ mass.}
\end{figure}

We now turn to discuss the $Z'\to gg\gamma$ decay. In Table \ref{TABLE2}, the contribution of the SM quarks to $\Gamma(Z'\to gg\gamma)$ is shown for $m_{Z'}=500$ GeV  and $m_{Z'}=1.5$ TeV. These values for the $Z'$ mass are consistent with the corresponding bounds given in the literature~\cite{ZpMass}. It can be appreciated from these tables that the main contribution arises from the third family, but that those contributions coming from the first and second families are also relevant, as they are lower than that of the third family by less than one order of magnitude. It is also important to notice that the interference effect among families is in magnitude as important as the contribution of the third family. This effect is destructive for relatively light $Z'$ gauge boson, but it is constructive for a heavier $Z'$ boson. As it can be appreciated from these tables, the SM quarks contribution to the decay width for $m_{Z'}=1.5$ TeV is almost one order of magnitude larger than for $m_{Z'}=500$ GeV. On the other hand, the contribution of the exotic quarks to $\Gamma(Z'\to gg\gamma)$ is shown in Table \ref{TABLE3} for the scenario $\{m_{Z'}=1.5\; \mathrm{TeV}, m_Q=m_{D,S,T}=700\; \mathrm{GeV} \}$. It can be appreciated from this table that the exotic quark contributions are about of one order of magnitude larger than those induced by the SM quarks. It is also interesting to notice that the $T$ quark interferes constructively with the other two exotic quarks, in contrast with the case of the corresponding decay into three gluons, in which this interference effect is of destructive nature~\cite{OP}. This effect is due to the fact that the amplitude is proportional to $Q_{Q_i}$ and $Q_T>0$, but $Q_{D,S}<0$. This arises as a consequence of the fact that $T$ appears as a component of an antitriplet of $SU_L(3)$, whereas $D$ and $S$ arise as components of triplets of the same group. As it shown in Table VIII of Ref.~\cite{OP}, the destructive nature of the interference effect considerably reduces the contribution of the exotic quarks to the $Z'$ decay into three gluons. In our case, the constructive nature of this effect leads to a contribution to $\Gamma(Z'\to gg\gamma)$ that is more than one order of magnitude larger than that associated with $\Gamma(Z'\to ggg)$. This behavior can clearly be appreciated from Fig.~\ref{Graphic}, where the contributions to $\mathrm{Br}(Z'\to gg\gamma)$ of standard and exotic quarks are shown separately.

Using for the total decay width of the $Z'$ boson the results given in Ref.~\cite{Zp331}, the branching ratio for the scenario $\{m_{Z'}=m_Q=m_{D,S,T}=500\; \mathrm{GeV}\}$ is given by
\begin{equation}
\mathrm{Br}(Z'\to gg\gamma)=1.53\times 10^{-6}\; ,
\end{equation}
which is determined essentially by the SM quarks. On the other hand, the corresponding branching ratio for the scenario $\{m_{Z'}=1.5\; \mathrm{TeV}, m_Q=m_{D,S,T}=700\; \mathrm{GeV} \}$ is given by
\begin{equation}
\mathrm{Br}(Z'\to gg\gamma)=1.06\times 10^{-4}\;.
\end{equation}
In this case, the contribution induced by the SM quarks is marginal.

\begin{table}[!h]
 \caption{\label{TABLE2} Family contribution to the $\Gamma(Z'\to gg\gamma)$ decay for $m_{Z'}=500\; \mathrm{GeV}$ and $m_{Z'}=1.5\; \mathrm{TeV}$. Here $\Gamma^I$ represent the interference effect induced by the three families into the width decay.}
\begin{ruledtabular}
 \begin{tabular}{|c|c|c|c|c|c|c|}
         & $m_{Z'}=500\,\, \mathrm{GeV}$ & $m_{Z'}=1.5\,\, \mathrm{TeV}$ & $m_{Z'}=500\,\, \mathrm{GeV}$ & $m_{Z'}=1.5\,\, \mathrm{TeV}$ & $m_{Z'}=500\,\, \mathrm{GeV}$ & $m_{Z'}=1.5\,\, \mathrm{TeV}$ \\
  Family & $\Gamma$ [GeV] & $\Gamma$ [GeV] & $\Gamma^I$ [GeV] & $\Gamma^I$ [GeV] & $\Gamma_{q_i}$ [GeV] & $\Gamma_{q_i}$ [GeV] \\
  \hline
  $u,d$ & $1.13\times 10^{-4}$ & $7.01\times 10^{-4}$ & - & - & - & - \\
 \hline
  $c,s$ & $1.12\times 10^{-4}$ & $6.92\times 10^{-4}$ & - & - & - & - \\
  \hline
  $t,b$ & $3.64\times 10^{-4}$ & $3.06\times 10^{-3}$ & - & - & - & - \\
  \hline
  Total & $5.89\times 10^{-4}$ & $4.45\times 10^{-3}$ & $-3.95\times 10^{-4}$ & $1.96\times 10^{-3}$ & $1.94\times 10^{-4}$ & $6.41\times 10^{-3}$
\end{tabular}
\end{ruledtabular}
\end{table}

\begin{table}
\caption{\label{TABLE3} Contribution of the exotic quarks to  $\Gamma(Z'\to gg\gamma)$ in the scenario $\{m_{Z'}=1.5\; \mathrm{TeV}, m_Q=m_{D,S,T}=700\; \mathrm{GeV} \}$. Interference effects are also shown.}
\begin{ruledtabular}
\begin{center}
 \begin{tabular}{|l|l|}
  Quark & $\Gamma_{Q_i}$ [GeV] \\
  \hline
  $D$   & $5.73\times 10^{-3}$ \\
  \hline
  $S$   & $5.73\times 10^{-3}$ \\
  \hline
  $T$   & $1.83\times 10^{-2}$ \\
  \hline
  $D-S$ & $1.13\times 10^{-2}$ \\
  \hline
  $D-T$ & $2.04\times 10^{-2}$ \\
  \hline
  $S-T$ & $2.04\times 10^{-2}$
\end{tabular}
\end{center}
\end{ruledtabular}
\end{table}

In conclusion, in this paper the rare $Z\to gg\gamma$ and $Z'\to gg\gamma$ decays were studied in the context of the minimal $331$ model. Numerical results for the corresponding decay widths and branching ratios were obtained from exact analytical expressions given in Ref.~\cite{OP}. The relative importance of the contributions induced by the known quarks and the exotic ones predicted by the model was analyzed. It was found that the $Z\to gg\gamma$ decay is dominated by the first two families, being insensitive to the presence of exotic quarks. As to the $Z'\to gg\gamma$ decay, it was found that this process is very sensitive to the exotic quark presence and its decay width can be more than one order of magnitude larger than that associated with the $Z'\to ggg$ decay.

\acknowledgments{We acknowledge financial support from CONACYT and
SNI (M\' exico).}

\end{document}